\documentclass[aps,prb,twocolumn,showpacs]{revtex4}

\usepackage{graphicx} 

\begin{document}
\topmargin -0.25in
\leftmargin -0.2in
\rightmargin 0.2in

\title{Para to Ortho transition of metallic dimers on Si(001)}

\author{Bikash C Gupta} 
\author{Inder P Batra}\email{ipbatra@uic.edu} 

\affiliation{Department of Physics, University 
of Illinois at Chicago, 845 West Taylor Street, Chicago, IL 60607-7059}

\date{\today}

\begin{abstract}

Extensive electronic structure calculations are performed to obtain
the stable geometries of metals like Al, Ga and In on the Si(001) 
surface at 0.5 ML and 1 ML coverages. Our results coupled with 
previous theoretical findings explain the recent experimental data 
in a comprehensive fashion. At low coverages, as shown by previous 
works, `Para' dimers give the lowest energy structure. With increasing
coverage beyond 0.5 ML, `Ortho' dimers become part of low energy 
configurations leading toward a `Para' to `Ortho' transition at 1 ML
coverage. For In mixed staggered dimers (`Ortho' and `Para') give the 
lowest energy configuration. For Ga, mixed dimers are non-staggered, 
while for Al `Para' to `Ortho' transition of dimers is complete.  
Thus at intermediate coverages between 0.5 and 1 ML, the `Ortho' and 
`Para' dimers may coexist on the surface. Consequently, 
this may be an explanation of the fact that the experimental 
observations can be successfully interpreted using either orientation. 
A supported zigzag structure at 0.5 ML, which resembles ${\rm (CH)_x}$, 
does not undergo a dimerization transition, and hence stays semi-metallic. 
Also, unlike ${\rm (CH)_x}$ the soliton formation is ruled out for this 
structure.
\end{abstract}

 \maketitle

\maketitle

\section{Introduction}

The study of metals on semiconductors dates back to the nineteenth 
century and has seen a vigorous recent revival due to tremendous 
interest in Nanotechnology. Our ability to manipulate atoms,
placing them at will on different surface sites to create exotic 
artificial structures, has led to further investigations \cite{dow} of
electronic and transport properties of free and supported nanowires.
The placement of trivalent atoms (Al, Ga and In) on Si(001) can lead
to the formation of low-dimensional structures, exhibiting
significant new electronic properties.

One can easily compute the electronic properties of free standing
nanowires. For this, one must first compute the total energy and
determine the possible stable structures. Such calculations have
indeed been carried out for nanowires \cite{por,por1,tor,tak,hak,sen,ind1} 
consisting of a wide
variety of atoms, {\em e.g.}, K, Al, Cu, Ni, Au and Si. A general
finding is that a zigzag structure in the form of an equilateral
triangle is the most stable \cite{por,por1,sen,ind1}. 
This can be understood as arising 
primarily due to the maximization of coordination number for each
atom in a quasi 1D structure. Another structure which also has a
local minimum, but not terribly stable, is a wide angle isosceles
triangle which somehow is reminiscent of the bulk environment.
For example, Si which is a four fold coordinated in the bulk
(tetrahedral angle $\sim 109^{\rm o}$) shows \cite{ind1} a local minimum
at an angle of $\sim 117^{\rm o}$. In general, free standing 
nanowires tend to be metallic (have bands crossing the Fermi level)
but these nanowires in practice are to be supported. Silicon is the most 
widely used substrate for practical applications and the low
index surfaces, Si(001) is the surface of choice. With the downward
spiral toward nano devices, it is desirable to study the 
metallic properties at the lowest possible coverages. It is in 
this context that the study of metals like Al, Ga and In at 
submonolayer coverages on Si(001) take on the added importance.
The interaction of metal nanowires with substrate can significantly
alter the electronic properties.

Low-energy-electron-diffraction (LEED), Auger electron spectroscopy 
(AES) and Scanning tunneling microscopy (STM) studies have provided 
much informations about the interaction of Al, Ga and In overlayers 
on the Si(001) surface at different coverages and at different 
temperatures \cite{ide,sakm,bou,knl,nog1,nog2,bas1,bas2,sak1,sak2,sak3,sak4}.

Ide et al. \cite{ide} observed 2$\times$2, 2$\times$3, 4$\times$5,  
and 1$\times$7 structures depending on the coverages less than 1 
monolayer (ML)  
of Al on Si(001) and the substrate temperature in their LEED and AES 
experiments. However, the 2$\times$2 and 2$\times$3 phases were observed 
around 0.5 and 1/3 ML coverages respectively at low temperatures.   

Sakamoto and Kawanami \cite{sakm} performed Reflection high energy electron 
diffraction experiments and established the existence of phases
with various symmetries (2 $\times$ 3, 2 $\times$5, 2$\times$2 and
8$\times$1) for Ga coverages less than 1 ML and at temperatures
$\sim$ 350$^{\rm o}$ C. Later, Bourguignon et al. 
\cite{bou} examined the evolution of 1ML of Ga on Si(001) 
with LEED and observed 
all the above mentioned phases and in addition they also observed 
1$\times$2 phase at 1 ML coverage. Nogami et al. \cite{nog1,nog2} 
did STM studies 
and observed 2$\times$2 at and below 0.5 ML coverage of Al, Ga 
and In on Si(001). Knall et al. \cite{knl} performed experiment 
for In adsorption
on Si(001) using RHED, LEED, AES and STM and observed transition
from 2$\times$2 to 2$\times$1 structure of In as coverage 
increased.  All the experimental results 
\cite{ide,sakm,bou,knl,nog1,nog2,bas1,bas2,sak1,sak2,sak3,sak4} 
point to the fact that at low 
temperatures Al, Ga and In form 2$\times$2 structure on the Si(001)
surface at 0.5 ML coverage. 

At higher metal coverage, Nagomi et al. \cite{nog1} observed 3D 
island formation of Al on Si(001) around 1 ML coverage while ordered 
structures were observed for Ga and In on Si(001) at 1 ML coverage 
without any island formation \cite{bou,knl}.  In particular, a
1$\times$2 structure was observed for Ga, and 2$\times$2, 
2$\times$1 structures were observed for In on Si(001). Since 
Ga does not form islands at 1ML coverage and it has been argued
that Al forming islands at 1 ML coverage is unusual, we
present a complete theoretical study for Al/Si(001) at 1 ML coverage.

Note that several experimental results were explained 
\cite{ide,bou,nog1,nog2,bas1,bas2,ind} in terms of 
metal dimers oriented parallel to the Si dimer rows {\em i.e.} metal 
dimers were assumed to be directed perpendicular to the Si dimers 
in the reconstructed surface. This orientation of metal 
dimers is named as `Ortho' dimers.
Electronic structural calculations were done by Batra \cite{ind} 
for Al on Si(001) at 0.5 ML coverage and  he explained the 
experimentally observed patterns in
terms of these `Ortho' dimers. Later, Northrup et al. \cite{joh} did  
theoretical calculations on the same system and showed that the 
orientation of metal dimers parallel to the Si dimers was  
most favorable. These dimers  orientated parallel to 
the Si dimers are called as `Para' dimers.  A series of 
experiments \cite{sak1,sak2,sak3} performed recently for the 
structural configuration of metal dimers on the Si(001) surface indeed
confirm the findings of Northrup et al. \cite{joh} at 0.5 ML 
coverage. To the best of our knowledge, there is hardly any 
detailed theoretical study at higher coverages (1 ML) for these
systems.

In this paper, we put significant effort to perform 
zero temperature electronic structural calculations for both 
0.5 ML and 1 ML coverages of
Al, Ga and In on the Si(001) surface and compare our results with the 
experimental and theoretical findings where available. We
attempt to put the work in perspective and fill the holes 
by presenting new results.
In particular, we find that our calculations at 0.5 ML give results 
in accord with Northrup's predictions and also with the recent 
experimental results obtained by Sakama et al \cite{sak1,sak2,sak3}. 
More significantly, 
we obtain an orientational transition (`Para' to `Ortho') for Al 
dimers as we go from 0.5 ML to 1 ML coverage. Our calculations 
also reveal that the reconstruction of Si(001) is completely lifted 
at 1 ML coverage of Al. 
We find that In and Ga interacts differently with the surface; for 
these metals  
the reconstruction is not lifted at 1 ML coverage. One finds the
staggered mixed (`Ortho' and `Para') dimers to be the lowest energy 
structure for In while the mixed dimers are non-staggered for Ga.
The important conclusion is that the `Ortho' dimers, which are 
energetically unfavorable at 0.5 ML coverage, become viable at 
higher coverage facilitating `Para' to `Ortho' transition as
a function of coverage.
Also, a previously reported semi-metallic phase of supported zigzag
Al nanowire is found to be stable against a  dimerizing Peierls 
transition unlike ${\rm (CH)_x}$. The soliton \cite{sol} 
formation in the supported nanowire is argued to be improbable. 

The paper is organized as follows. The calculational parameters are
given in sec. II and the results and discussions are presented
in sec. III followed by a summary of our findings in sec. IV.

\section{Method}

First principle total energy calculations were carried out within 
density functional theory at zero temperature using the VASP code  
\cite{kres}. The wave functions are expressed by plane waves with 
the cutoff energy $|k + G|^2 \le 250$ eV. The Brillouin Zone (BZ) 
integrations are performed by using the Monkhorst-Pack scheme with 
4$\times$4$\times$1 $k$-point meshes for 2$\times$2 primitive cells. 
Ions are 
represented by ultra-soft Vanderbilt type pseudopotentials and results 
for fully relaxed atomic structures are obtained using the 
generalized gradient approximation (GGA). The preconditioned 
conjugate gradient method is used for the wave function optimization 
and the conjugate gradient method for ionic relaxation.

The Si(001) surface is represented by a repeated slab geometry. 
Each slab contains five Si atomic planes with Hydrogen atoms
passivating the Si atoms at the bottom of the slab. Consecutive
slabs are separated by a vacuum space of 9 \AA. The Si atoms on 
the top four layer of the slab are allowed to relax while those 
in the bottom layer of the slab and the passivating Hydrogen 
atoms are kept fixed to simulate the bulk like termination. 
The convergence with respect to the energy cutoff and the number of
$k$ points for similar structures has been established \cite{sen1}.

\section{results and discussions}

We have performed total energy calculations using the above parameters 
for both the ideal and the reconstructed Si (001) surfaces. For the
ideal case, the Si atoms on the surface are arranged in a square
pattern of side 3.84 \AA. For the reconstructed surface the Si atoms 
rearrange in such a way that the top layer Si atoms form dimer rows 
along the $y$ [1$\bar {\rm 1}$0] direction but the atoms move along
the $x$ [110] direction to form dimers. Here we consider 2$\times$2 
supercell for our calculations.  At the 0.5 ML coverage of Al, Ga 
and In we start only with the reconstructed surface and at 1 ML 
coverage we start with both the ideal and reconstructed surfaces.

\subsection{0.5 ML of coverage}

Here we consider various structural arrangements of metal atoms on 
the Si(001) surface. The $2 \times 2$ primitive cell contains two metal
atoms which corresponds to 0.5 ML coverage. The results 
for Al/Si(001) are elaborated here and the results for Ga and In 
are briefly discussed. Two metal atoms in the cell may be placed 
in various ways on the surface. We consider several 
configurations to identify energetically the most favorable 
configuration. The configurations considered here are shown in the 
Fig. 1. 

\begin{figure}[ht]
\includegraphics[scale=0.35]{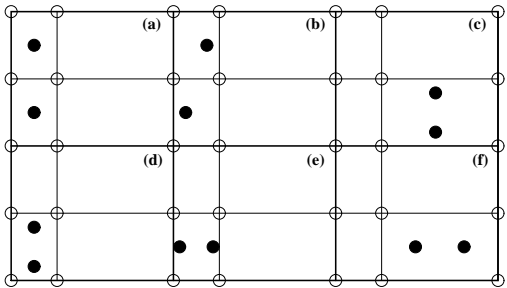}
\caption{Top layer of reconstructed Si(001) surface. The 2$\times$2 cells
closed by thick borders are shown for different configurations of metal atoms
at 0.5 ML coverage. The unfilled circles represent the top layer Si atoms
and the filled circles represent the metal atoms.}
\label{fig:fig1}
\end{figure}

The metallic dimers configurations on reconstructed Si(001) shown 
in Figs. 1(a), 1(b) and 1(c) were considered by 
Batra \cite{ind}. He found that the configuration
in Fig. 1(b) was the most stable one followed by the configuration in
Fig. 1(c). The configurations in Figs. 1(c) and 1(f) were considered 
by Northrup et al \cite{joh} and they concluded that configuration 
in 1(f) was energetically the most favorable one. The arrangement 
in 1(c) is referred to as the `Ortho' orientation of metal dimers and 
that in 1(f) is the `Para' orientation of metal dimers as described
earlier by the Northrup et al \cite{joh}. In addition we consider two 
more configurations and they are shown in Figs. 1(d) and 1(e) 
respectively. The relative energies and the distance between the neighbor 
atoms for all the configurations are given in table I. All the 
energies are measured with respect to the reference energy which at 
0.5 ML coverage of metal atoms correspond to the configuration shown 
in Fig. 2(a). For calculating the reference energy, all the Si atoms 
are held fixed and the metal 
atoms are allowed to relax along the [001] direction. The reference 
energies for the cases of Al, Ga, and In are -131.625 eV, -130.714 
eV, and -130.255 eV. respectively. It is clear from table I that 
the configuration shown in 1(f) {\em i.e.}, Al dimer placed between the 
Si dimer rows with `Para' orientation  has the lowest energy and hence 
the most favorable configuration. Our results support the 
calculations first made by Northrup et al \cite{joh} and later by
Brocks et al\cite{car}. We also observe that the next most favorable 
configuration is that shown in 1(e) which also 
has `Para' orientation of Al dimer but the dimer is placed within a 
Si dimer row. However, the energy difference between these two 
configurations with `Para' orientation of metal dimers is $\sim$ 1 eV. 
This is because the dimer within the Si row [1(e)] has to reside at much 
higher position compared to that placed between the dimer rows 
[1(f)] and hence has a weaker binding. 
We also note that the energy difference between the `Para' and `Ortho' 
orientation of dimers is even larger, the `Para' being more
stable $\sim$ 1.3 eV. 

\begin{figure}[ht]
\includegraphics[scale=0.35]{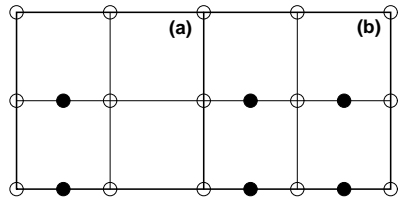}
\caption{The configurations (a) and (b) are considered to calculate the
reference energies at 0.5 and 1 ML metal coverages respectively. 
The unfilled circles and the filled circles represent the top layer 
Si atoms and the metal atoms respectively. Here all the Si atoms are 
held fixed and the metal atoms are allowed to
relax along the (001) direction.} 
\label{fig:fig2}
\end{figure}

To calculate the potential energy variation of the system  as a function 
of the Al dimer orientation, we place the Al dimer between the
Si dimer rows at a fixed height and rotate about the [001] direction. 
The result is shown in Fig. 3.  This shows two minima at `Ortho' and 
`Para' orientations respectively and there is a small barrier as one goes 
from `Ortho' to `Para' orientation.  Therefore, the Al dimers 
take `Para' orientation which is strongly more favorable compared to the 
`Ortho' orientation.  This result also agrees with the observation in 
a recent experiment by Sakama et al \cite{sak1}. We note that though 
the Si dimers are stretched by 0.26 \AA~ the reconstruction of the Si 
surface is not lifted at 0.5 ML of coverage.

We have also done similar calculations for Ga and In adsorption on
the reconstructed Si(001) surface and found that the `Para' orientation
of Ga and In dimers are also favored with relative energies -3.29 eV
and -2.90 eV respectively. The results for Ga and In also 
agree with recent experimental results obtained by Sakama et al 
\cite{sak2,sak3}. The surface pattern at 0.5 ML coverage is 
2$\times$2. It is also clear that all the patterns observed at low 
temperatures and below 0.5 ML coverage of Al, Ga and In on the 
reconstructed Si(001) can be explained by the `Para' dimers as 
proposed by Northrup \cite{joh}, but contrary to the earlier proposal 
by Batra \cite{ind}.  

\begin{figure}[ht]
\includegraphics[scale=0.35]{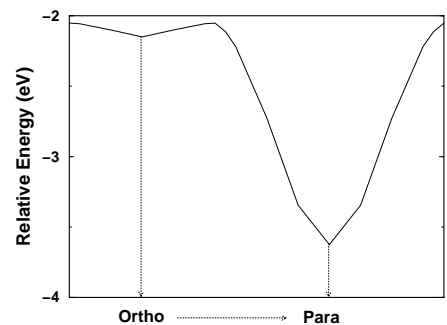}
\caption{The variation of energy of a 2$\times$2 reconstructed cell as the
Al dimer is rotated from `Ortho' to `Para' orientation. The Al dimer is
placed between the Si dimer rows of the reconstructed Si(001) surface.}
\label{fig:fig3} 
\end{figure}

One focus of Batra's work \cite{ind}, however, was to find a stable
metallic configuration for Al. He reported a zigzag nano-structure 
shown in Fig. 1(b) which he labeled as ${\rm R_x(H1,H2)}$, as being
stable with a local minimum in energy. Our present calculations 
support this finding except that we get $\Delta x = \pm 0.6$ \AA~ 
in contrast to Batra's value of $\pm 0.8$ \AA. This overlayer structure 
of Al atoms
is reminiscent of a uniform bond ${\rm (CH)_x}$ structure
with the important difference that there is no direct Al-Al
bond. The nearest neighbor Al-Al distance ($\sim$ 4 \AA)~ in the
overlayer is much longer than the bulk Al-Al distance 
($\sim$ 2.8 \AA) and they ``see" each other only through the
delocalized electrons in the $p_z$-orbitals. In ${\rm (CH)_x}$,
C-C atoms are not only coupled through $\pi$-electrons but
they also form strong uniform $\sigma$-bonds leading to a
semi-metallic structure. This structure becomes semi-conducting
upon undergoing a bond alternation Peierls like transition to
Double (D), Single (S), DS, DS ... repeating pattern of bonds. 
It supports the soliton \cite{sol} formation by creating a defective 
structure of the type .. DS, DS, SD, SD .. by having two 
single bonds adjacent to each other. The Al overlayer 
structure\cite{ind2}
has two electrons of each Al atom tied to a surface Si
atom in an ${\rm AlX_2}$ like fashion while the third
electron of Al is in a free electron like state along the
nanowire ($y$-direction). Here X represents a surface Si
atom which nominally has a single electron in a dangling
$p_z$-orbital available for bonding with Al. Interatomic 
distance between Al and surface Si atoms, d(Al-X) $\approx$
2.4 \AA,~ implies a strong bond. The delocalized $p_z$ electron 
(one per Al atom) creates a one dimensional semi-metallic
system. Naturally, under such situation metallic wires may be realized 
and it will be conducting under small bias conditions. 
But then the standard Peierls distortion may come into play
depending upon the strength of the surface bonds. 
Within the error of our calculations (estimated at $\sim$
7 meV) we did not find any lowering of the total energy upon 
dimerization, unlike $(CH)_{\rm x}$. In fact we noted
that a dimerization of Al atoms in the zigzag structure by 
an amount $\Delta y = \pm 0.15$ \AA~ raised the total energy 
by $\sim$ 0.1 eV. Thus the surface bonds formed by supported
nanowires can lock the metallic structure in place.
There is no possibility of a soliton 
formation in the Al overlayer structure because of the large Al-Al
distance. However, it will be 
interesting to look for soliton behavior in other overlayers.

\subsection{1 ML of coverage}

We consider the adsorption of four metal atoms on the Si(001)
surface of a 2$\times$2 unit cell which corresponds to 1 ML 
coverage.  Results for Al, Ga, and In on Si(001) are presented 
here. Various possible configurations considered  
are shown in Fig. 4. Here calculations are done for both the 
ideal and the reconstructed Si surfaces to find energetically 
the most favorable configuration. The 2$\times$2 unit 
cells are bordered by thick lines in Fig. 4. The configurations
considereed include Para-Para and Ortho-Ortho metal dimers on
reconstructed (Figs. 4(a) - 4(d)) and Ideal (Figs. 4(e) - 4(h))
on Si(001) surface. In addition, we also consider mixed dimers
(Ortho and Para) on reconstructed surfaces (Figs. 4(i) -  4(l)).
Mixed dimers on ideal Si surface were not stable and readily
caused the silicon surface to reconstruct. 

\begin{figure}[ht]
\includegraphics[scale=0.35]{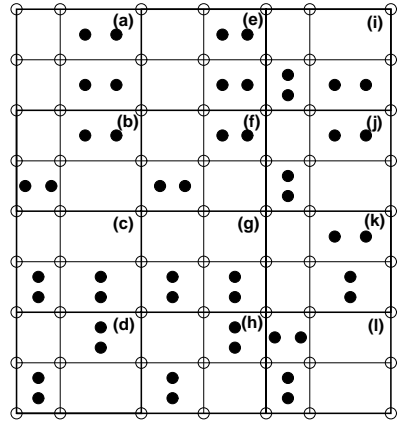}
\caption{Metal atoms on top layer of reconstructed and ideal Si(001) 
surface. The 2$\times$2 cells closed by thick borders are shown for 
different configurations metal atoms at 1 ML coverage. The unfilled 
circles represent the top layer Si atoms and the filled atoms represent 
the metal atoms (Al or Ga or In).} \label{fig:fig4} 
\end{figure}

The relative energies for the relaxed structures of Al, Ga and In 
on the reconstructed surfaces are given in table II for four possible 
configurations and the same on the ideal surfaces are shown in 
table III. The energies for the configurations with mixed dimers are 
given in table IV. The reference energies at 1 ML coverage of metal 
atoms are obtained for the configuration shown in Fig. 2(b) where all
Si atoms are held fixed the metal atoms are allowed to relax along
the (001) direction. The reference energies for the cases of Al, Ga 
and In are -137.963 eV, -136.485 eV and -136.044 eV respectively.

A general conclusion is that with increasing coverage beyond 0.5 ML,
`Ortho' dimers become part of the low energy configurations leading 
towards a `Para' to `Ortho' transition at 1 ML coverage. For In,
mixed staggered dimers (`Ortho' and `Para') give the lowest energy
configuration. For Ga, mixed dimers are non-staggered, while for
Al `Para' to `Ortho' transition of dimers is complete. This is 
an important conclusion because at or below 0.5 ML `Ortho' dimers
were not favorable. Also, the puzzle that `Ortho' and `Para' dimers
are both seen is resolved due to the preponderance of `Ortho'
dimers at increasing coverages. 

Let us first consider the structure consisting of Al dimers with
`Para' orientation as it was the most favorable structure 
at 0.5 ML coverage. Two configurations are possible with `Para' 
orientation of Al dimers: (i) the configurations shown in Figs. 4(a)  
and 4(e) on the reconstructed and ideal  surfaces respectively and 
(ii) the configurations shown in Figs. 4(b) and 4(f) on the 
reconstructed and ideal surfaces respectively. After complete relaxation, 
we note that  the final low energy structure  
is independent of the initial surface pattern (reconstructed
/ideal). When we start from an ideal surface, the Si atoms 
dimerize  and the low energy structure is the one with reconstructed 
Si surface. Same final structure is achieved when we start form the 
reconstructed surface.  Also note that the relaxed system with the 
staggered dimers configuration in 4(b) consisting of parallel 
dimers on the reconstructed surface is less favorable compared to 
that on the same reconstructed surface with the configuration in 4(a). 
Comparing 4(b) and 4(f) we find that the final energies of the systems 
differ by a large amount: staggered dimers on the ideal surface
are least favorable.

Next we consider the configurations in  4(c) and 4(d). These structures 
are obtained by just rotating the Al dimers in 4(a) and 4(b) by 90 
degree ({\em i.e.}, dimers now take `Ortho' orientation). From
table II it is clear that these configurations are more favorable
compared to those in 4(a) and 4(b). We find that after complete
relaxation, the total energies and the structures of these two 
configurations become independent of the initial surface pattern.

In addition, we consider the configurations consisting of mixed
dimers (`Ortho' and `Para') as shown in Figs. 4(i) - 4(l). Similar
configurations on ideal surface are unstable, the Si surface 
reconstructs itself and eventually leads to the configurations
4(i), 4(j) and 4(k). Again for Al, the configurations 4(j) and 4(l) are
unstable. From table IV we see that the configuration 
with mixed Al dimers shown in Fig. 4(i) has a deep local minumum. This 
configuration is more favorable compared to the configurations
consisting of only `Para' orientation of Al dimers (i.e., 
Figs. 4(a) and 4(f)). 
 
However, comparing the energies for configurations 4(i) and 4(g)
we find that the lowest energy structure is the one shown in 
Fig. 4(g) where the reconstruction is completely lifted. 
All the Si bonds are saturated here.  Since the reconstruction is 
completely lifted at 1 ML coverage, we conclude that the breaking 
of Si dimer bonds is initiated just beyond 0.5 ML of coverage. 
Most interestingly we note that an orientational transition for 
the Al dimers takes place as one goes from 0.5 ML coverage to 1 ML 
coverage: The Al dimers change their orientation from `Para' to `Ortho' 
after breaking the Si dimer bonds. This is also supported by the charge 
density plot for the lowest energy configurations at 0.5 and 1 ML 
coverage as shown in Figs. 5(a) and  5(b) respectively.

\begin{figure}[ht]
\includegraphics[scale=0.7]{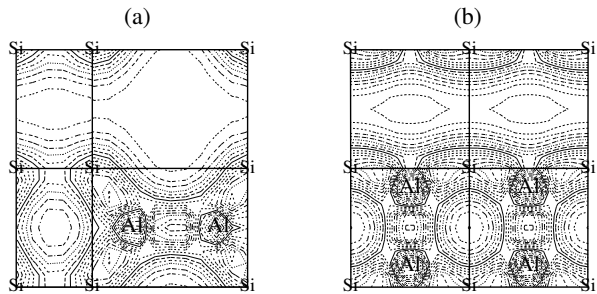}
\caption{The charge density plot for: (a) the lowest energy structure 
at 0.5 ML coverage, (b) the lowest energy structure at 1 ML coverage. 
The charge density plane coincides with the Al layer.} 
\label{fig:fig5} 
\end{figure}

Earlier, it was argued that at 0.5 ML coverage all Si bonds 
are saturated and therefore beyond 0.5 ML coverage Al atoms 
may reside on top of each other and form 3D island. One experimental 
study observed the 3D island formation beyond 0.5 ML of Al coverage 
\cite{nog1}. Our calculation, however, suggests that for Al an  
orientational transition can take place at 1 ML coverage and the 
surface structure becomes  regular having  a 1$\times2$ pattern 
where all bonds of Si and Al are saturated due to removal of 
the reconstruction of the Si surface.  This leads us to speculate 
that further experiments on Al may seek out 1$\times$2 pattern at 1 ML 
coverage. Furthermore, at coverages between 0.5 ML and 1 ML  
both the `Para' and `Ortho' Al dimers may coexist. This also 
clarifies why experiments at times have reported both `Para' 
\& `Ortho' dimers because precise coverage is a difficult parameter 
to quantify. 

\begin{figure}[ht]
\includegraphics[scale=0.5]{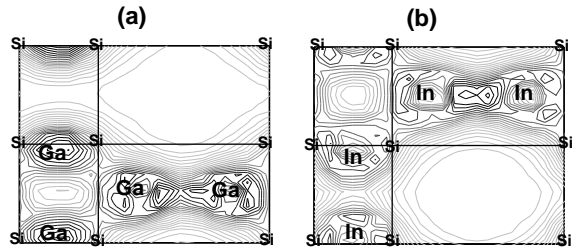}
\caption{The charge density plot for: (a) the lowest energy structure
at at 1 ML coverage of Ga on Si(001), (b) the lowest energy structure
at at 1 ML coverage of In on Si(001)} 
\label{fig:fig7} 
\end{figure}

Now we consider the Ga adsorption at 1 ML coverage. If 
we start with the reconstructed surface we find that the 
configuration in Fig. 4(i) with mixed dimers leads to the lowest
energy structure where the reconstruction of the Si surface is
not lifted during the relaxation process. The next best 
configuration is shown in Fig. 4(g) and consists of only `Ortho'
dimers. The structure remains ideal even after complete ionic
relaxation. The configurations in Figs. 4(j) and 4(l) are
unstable. Comparing final energies for various configurations 
on ideal and reconstructed surfaces we observe that the 
configuration in 4(i) is energetically the most favorable 
structure. Experiments indicate observing the 1$\times$2 
structure shown in 4(g) \cite{bou}. Here the reconstruction of 
the Si surface is lifted, Ga dimers prefer `Ortho' orientation 
leading to a 1$\times$2 patten. Our calculations suggest
that mixed dimers (Fig. 4(i)) and `Ortho' dimers (Fig. 4(g))
as two contenders for the energy favored structures.  
 
For the case of In we find that energetically the most 
favorable configuration is given in Fig. 4(j). This configuration 
corresponds to a 2$\times$2 surface pattern. The configurations
in Figs. 4(i) and 4(l) are unstable. More significantly, we
note that the reconstruction of the surface is not lifted.
We also find that the configurations consisting of `Para'
dimers (i.e., Figs. 4(a) and a(b)) are degenerate and 
energetically next favorable structures. The configurations
in Figs. 4(a) and 4(b) correspond to the 2$\times$1 and 
2$\times$2 surface structures respectively. Experimental
results indicate the existence of both the 2$\times$2 and 
2$\times$1 surface structures near 1 ML coverage of In.
The most and the next favorable structures from our 
calculations are consistent with the experimentally 
observed 2$\times$2 and 2$\times$1 surface structures 
respectively.

To understand the bonding of Ga and In atoms on Si(001),
the charge density distributions are plotted on the plane of 
metal atoms for the most favorable configurations at 1ML coverage 
of Ga and In and they are shown in Figs. 6(a) and 6(b) 
respectively. We observe that to a first approximation
metal atoms display strong intra-plane bondings amongst
themselves leaving the substrate in reconstructed state. 
The charge distribution in Fig. 6(b)
also helps us to understand why the configuration 4(i) is 
unstable for In. In being larger in size, the In atoms
in `Ortho' orientation of configuration 4(i) are pushed
away from each other to have a suitable coupling among In
atoms and thus favoring the configuration 4(j) with (mixed) 
staggered dimers.

It is worth mentioning that one usually ends up with semi-metallic
or non-metallic nanowires when supported on substrates as we 
saw above for the 
group III metals on Si(001). Very recently, there have been several 
important studies \cite{him1,him2,him3,him4,him5} of monovalent
Au chains on various high indexed Si surfaces dealing
with metallicity, Peierls distortion, charge density instability,
fractional band fillings, and possibility of a spin charge 
separation due to the Luttinger-liquid phase. In fact, 
Himpsel and his group showed that Au nanowire supported on Si(553) 
is stable and it shows metallic behavior \cite{him1,him5}.

\section{summary}

We have done extensive total energy calculations for Al, Ga and 
In adsorption on the Si(001) surface at 0.5 and 1 ML coverages.
The `Para' orientation of dimers is the most stable 
orientation at 0.5 ML coverage for Al, Ga and In, which 
supports the results of Northrup et al and also agrees with the 
recent experimental observations. All the patterns observed in 
various experiments at low temperature and below 0.5 ML coverages 
can be explained by the `Para' orientation of metal dimers. 
An interesting zigzag semi-metallic structure similar to ${\rm (CH)_x}$
is formed at 0.5 ML but is not the lowest energy structure.
Soliton formation in overlayers is a fundamental topic for 
further study.

At 1 ML coverage a tendency towards the `Ortho' orientation of 
dimers is noted. 
Thus we predict an orientational transition from `Para' 
to `Ortho' as we go from 0.5 ML to 1 ML coverage for Al and possibly 
Ga on the Si surface. In addition, the reconstruction of the Si 
surface is completely lifted at 1 ML coverage  of Al. These 
results lead us to conclude that the surface patterns at 0.5 ML
coverage can be explained in terms of `Para' orientation
of metal dimers and those  at intermediate coverages between 0.5 
and 1 ML may be explained by the co-existence of `Para' and `Ortho' 
orientation of dimers. For In mixed `Para' and `Ortho' dimers give
the lowest energy configuration that exhibit 2$\times$2 surface 
pattern and the transition is not complete.

\begin{table*}
\caption{Relative energies of the relaxed 2$\times$2 cell 
at 0.5 ML coverage of Al on the {\bf reconstructed} Si(001) surface for 
various configurations. Various bond lengths are also given. The
reference energy for this case is -131.625 eV corresponding to the
configuration in Fig. 2(a).}
\label{table:table1}
\begin{ruledtabular}
\begin{tabular}{|c|c|c|c|c|c|}
\hline
{\bf Configuration in figure 1}& {\bf Energy (eV)}& {\bf d(Si-Si)}&
{\bf d(Si-Al)}& {\bf d(Al-Al)}& {\bf height} \\
\hline
1(a)& -1.892 &2.52 &2.67 &3.84 &1.36  \\
\hline
1(b)& -1.951 &2.46 &2.50 &4.01 &1.40  \\
\hline
1(c)& -2.216 &2.78 &2.62 &2.64 &0.73  \\
\hline
1(d)&-2.153 &2.59 &2.39 &2.60 &1.92  \\
\hline
1(e)&-2.499 &2.45 &2.47 &2.62 &1.55  \\
\hline
1(f)&-3.482 &2.44 &2.48 &2.67 &1.10  \\
\hline
\end{tabular}
\end{ruledtabular}
\end{table*}

\begin{table*}
\caption{Relative energies of a relaxed 2$\times$2 cell at 1ML coverage 
of Al, Ga and In on a {\bf reconstructed} Si(001) surface for four different 
configurations with either `Ortho' or `Para' dimers. "R" and "I" 
implies that the Si(001) surface becomes 
"Reconstructed" and "Ideal" respectively after complete ionic relaxation.
The reference energies for the case of Al, Ga and In are -137.963 eV, 
-136.485 eV and -136.044 eV respectively corresponding to the configuration
in Fig. 2(b).}
\label{table:table2}
\begin{ruledtabular}
\begin{tabular}{|c|c|c|c|}
\hline
{\bf Configuration in figure 4}& {\bf Al: Energy (eV)}& {\bf Ga: Energy (eV)}&
{\bf In: Energy (eV)} \\
\hline
4(a)&-2.680 (R) &-2.528 (R) &-2.077 (R)  \\
\hline
4(b)&-2.371 (R) &-2.448 (R) &-2.077 (R)  \\
\hline
4(c)&-3.092 (I) &-2.148 (R) &-1.441 (R)  \\
\hline
4(d)&-2.868 (I) &-2.187 (R) &-1.441 (R) \\
\hline
\end{tabular}
\end{ruledtabular}
\end{table*}

\begin{table*}
\caption{Relative energies of a relaxed 2$\times$2 cell at 1ML coverage 
of Al, Ga and In on an {\bf ideal} Si(001) surface for four different 
configurations with either `Ortho' or `Para' dimers. 
"R" and "I" implies that the Si(001) surface becomes
"Reconstructed" and "Ideal" respectively after complete ionic relaxation.
The reference energies for the case of Al, Ga and In are -137.963 eV,
-136.485 eV and -136.044 eV respectively corresponding to the 
configuration in Fig. 2(b).}
\label{table:table3}
\begin{ruledtabular}
\begin{tabular}{|c|c|c|c|}
\hline
{\bf Configuration in figure 4}& {\bf Al: Energy (eV)}& {\bf Ga: Energy (eV)}&
{\bf In: Energy (eV)} \\
\hline
4(e)&-2.680 (R) &-2.528 (R) &-1.840 (R)  \\
\hline
4(f)&-0.339 (I) &-0.427 (I) &0.099 (I)  \\
\hline
4(g)&-3.092 (I) &-2.739 (I) &-0.389 (I)  \\
\hline
4(h)&-2.868 (I) &-2.631 (I) &-0.364 (I) \\
\hline
\end{tabular}
\end{ruledtabular}
\end{table*}

\begin{table*}
\caption{Relative energies of a relaxed 2$\times$2 cell at 1ML coverage 
of Al, Ga and In on a {\bf reconstructed} Si(001) surface for four different 
configurations consisting of mixed (`Ortho' and `Para') dimers.
"R" implies that the Si(001) surface becomes "Reconstructed" after complete
ionic relaxation. The configurations marked `Unstable' transforms to
one of other stable structure upon ionic relaxation. The reference energies 
for the case of Al, Ga and In are -137.963 eV, -136.485 eV and -136.044 eV 
respectively corresponding to the configuration in Fig. 2(b).}
\label{table:table3}
\begin{ruledtabular}
\begin{tabular}{|c|c|c|c|}
\hline
{\bf Configuration in figure 4}& {\bf Al: Energy (eV)}& {\bf Ga: Energy (eV)}&
{\bf In: Energy (eV)} \\
\hline
4(i)&-2.937 (R) &-2.995 (R) & Unstable   \\
\hline
4(j)& Unstable & Unstable  & -2.496 (R)  \\
\hline
4(k)&-2.667 (R) &-2.195 (R) &-1.886 (R)  \\
\hline
4(l)& Unstable & Unstable & Unstable \\
\hline
\end{tabular}
\end{ruledtabular}
\end{table*}

\end{document}